\documentclass[a4paper,12pt]{article}
\usepackage{cmap}
\usepackage[T2A]{fontenc}             
\usepackage[utf8]{inputenc}           
\usepackage[russian,english]{babel} 
\usepackage{amsmath}                  
\usepackage{amsfonts}
\usepackage{amssymb}
\usepackage{exscale}         
\usepackage[pdftex,unicode]{hyperref}
\hypersetup{
	colorlinks,%
	citecolor=blue,%
	filecolor=blue,%
	linkcolor=blue,%
	urlcolor=blue
}                                         
\usepackage[top=2cm, left=2cm, right=2cm, bottom=2cm]{geometry} 
\numberwithin{equation}{section}
\usepackage[style=gost-numeric,
backend=biber,
language=auto,
hyperref=auto,
autolang=other,
sorting=none
]{biblatex}
\begin{document}
\title{$ T $-matrix scattering elements for coulomb interaction systems.}
\author{R.F. Akhmetyanov\thanks{robertu@mail.ru} ,  E.S. Shikhovtseva  \\
	 \itshape Institute of Molecule and Crystal Physics - Subdivision of the Ufa \\ Federal Research Centre of the Russian Academy of Sciences \\ (IMCP UFRC RAS) 
	 \\  
	 \itshape (Prospekt Oktyabrya 151, Ufa, Russia, 450075) }
\date{}	
\maketitle         

\begin{abstract}
The paper derives the representation of the two-particle T-matrix scattering elements for the Coulomb interaction
with respect to special bases without expansion in terms of partial waves. The results obtained are applicable
to small-particle systems. The advantage of this expansion also arises in three-body problems when solving
the Faddeev equation for three-particle systems. The main problem in solving the Faddeev equation is the approximate
choice of approximation for the interaction potentials, at which the T-matrix scattering elements acquire a
separable form. However, even in this case the solution to the Faddeev equation does not always become practical
in view of the fact that the T-matrix elements themselves do not factor in the integral equations. Here we give the
results with the T-matrix elements represented in the basis, for which there is an addition theorem and hence the
integral Faddeev equations are reduced to a factored form. 
\end{abstract}
\begin{quote}
\textbf{Keywords:} Coulomb systems, scattering matrix, hypergeometric function.
\end{quote}

\newpage
\selectlanguage{russian}
\begin{center}
{\Large \textbf{$ T $-матричные элементы рассеяния для кулоновских систем взаимодействия.}} \\
\bigskip
Ахметьянов Р.Ф., Шиховцева Е.С \\
\medskip
\textit{ Институт физики молекул и кристаллов УФИЦ РАН, \\ 
Россия, 450075, г. Уфа, пр. Октября, 151 \\
E-mail: robertu@mail.ru}
\end{center}

\begin{quote}
 \textbf{Аннотация:} В работе содержится вывод представления двухчастичных $ T $-матричных элементов рассеяния для кулоновского взаимодействия по специальным базисам без разложения по парциальным волнам. Полученные результаты применимы к малочастичным системам. Преимущество данного разложения  возникает и в задачах трех тел при решении уравнения Фаддеева для трёхчастичных систем. Основной проблемой решении уравнения Фаддеева является приближенный выбор аппроксимации потенциалов взаимодействии, при котором  $ T $-матричные элементы рассеяния приобретали сепарабельный вид. Однако даже в таком случае решение уравнения Фаддеева не всегда становятся практичным в виду того, что входящие $ T $-матричные элементы в интегральные уравнения уже не факторизуются. Здесь мы представим результаты, в котором  $ T $-матричные элементы представляются в базисе, для которых существует теорема сложения  и вследствие чего интегральные уравнения Фаддеева приводятся к к факторизованному виду. \\

\textbf{Ключевые слова:} кулоновские системы, матрица рассеяния, гипергеометрическая функция. 
\end{quote}

\section{Введение.} 
Применение разложения обратно степенных потенциалов взаимодействия от трехмерных векторов по сферическим функциям широко используется в физических и математических задачах, обладающих сферической симметрии. Однако возможно особый интерес в физических и математических приложениях и задачах представляет не разделение по отдельности угловым и пространственным переменным, а разделение по полным векторам. Как было показано в \cite{AkhR_UNC} такое разделение существует для трехмерных двух векторов, и в конечных результатах угловые и пространственные переменные входят равноправно. К примеру, в задачах многих тел \cite{Dzibuti}, \cite{Shmidt_problem_tree_body}  появляется возможность не разделять отдельно гиперсферические угловые функции и решать отдельно систему по пространственным координатам, а решать общую систему по полным векторам. В работе рассматривается двухчастичная  $ T $-матрица в импульсном представлении, определяемая интегральным уравнением \cite{Shmidt_problem_tree_body}
\begin{equation} \label{1.1} 
\left< \mathbf{k}_2 \left| T(z) \right| \mathbf{k}_1 \right>=
\left< \mathbf{k}_2 \left|V \right| \mathbf{k}_1 \right>+
\int \!\! \frac{d\mathbf{p}}{(2\pi)^{\frac{3}{2}}}\,
\frac{\left< \mathbf{k}_2 \left|V \right| \mathbf{p} \right> \left< \mathbf{p} \left| T(z) \right| \mathbf{k}_1 \right> }{z-\dfrac{\hbar^2p^2}{2\mu}}, \,\,\,(z=E+i0) 
\end{equation}
здесь  $ E $-энергия относительного движения двух частиц,  $ \mu $-приведенная масса. Отметим, что амплитуда упругого рассеяния частиц выражается через  $ T $-матрицу как \cite{Shmidt_problem_tree_body}
\begin{equation*}
f(\mathbf{q}_2,\mathbf{q}_1)=-\frac{\mu}{2\pi\hbar^2}\left< \mathbf{q}_2 \left|T(E+i0) \right| \mathbf{q}_1 \right>
\end{equation*}
на энергетической поверхности  $ E=\dfrac{\hbar^2\,q_1^2}{2\mu}=\dfrac{\hbar^2q_2^2}{2\mu} $, где  $ \mathbf{q}_1 $ -- налетающий импульс,  $ \mathbf{q}_2 $ -- рассеянный. Для кулоновского поля  $ V(\mathbf{r})=\dfrac{\sigma\,\alpha}{r} $  потенциал взаимодействия в импульсном представлении есть как
\begin{gather} \label{1.2} 
\left< \mathbf{k} \left|V \right| \mathbf{p} \right>=
\int\!\! \frac{d\mathbf{r}}{(2\pi)^{\frac{3}{2}}} \, V(\mathbf{r}) e^{-i(\mathbf{k}-\mathbf{p})\mathbf{r}}
=\sigma\alpha\sqrt{\frac{2}{\pi}}\frac{1}{|\mathbf{k}-\mathbf{p}|^2}=\frac{\sigma\alpha}{\gamma^2}\,\sqrt{\frac{2}{\pi}}\, 
\left| \dfrac{\mathbf{k}}{\gamma}-\dfrac{\mathbf{p}}{\gamma} \right|^{-2} 
\\ 
\alpha=\left|  \frac{Z_1Z_2e^2}{4\pi\epsilon_0} \right| ,\,\, \sigma=\pm 1 \notag
\end{gather}
где в последнем выражении мы используем условие однородности функции,  $ \gamma $ --любое число которое можно задать в дальнейшем. Здесь $ \sigma =+1 $  соответствует потенциалу отталкивания двух зарядов $ Z_1e $  и $ Z_2e $ , а   $ \sigma =-1 $ потенциалу притяжения.

\section{Матричная формулировка.} 

Представим \eqref{1.2} в виде разложения из \cite{AkhR_UNC}, \cite{Robert_1711.07337} для трёхмерных векторов как (здесь и далее для сокращенной записи $ \sum\limits_{n,l,m}=\sum\limits_{n=0}^{\infty}\sum\limits_{l=0}^{\infty}\sum\limits_{m=-l}^{+l} $  )
\begin{equation} \label{2.1}
\left< \mathbf{k} \left|V \right| \mathbf{p} \right>=
\frac{\sigma\alpha}{2\gamma^{4}} (2\pi)^{\frac{3}{2}} \,
\sqrt{k^2+\gamma^2}\sqrt{p^2+\gamma^2}
\sum_{n,l,m}\upsilon_{n,l}
H_{n,l,m}\left( \frac{\mathbf{k}}{\gamma} \right)
H_{n,l,m}^{\ast}\left( \frac{\mathbf{p}}{\gamma} \right)
\end{equation}
\begin{equation*}
\upsilon_{n,l}=\frac{1}{n+l+1}
\end{equation*}
где функции $ H_{n,l,m}( \mathbf{k}) $  определяются в виде
\begin{equation} \label{2.2} 
H_{n,l,m}\left( \mathbf{k} \right)=\eta_{n,l}(k)\,Y_{l,m}(\hat{\mathbf{k}})
\end{equation}
здесь $ Y_{l,m}(\hat{\mathbf{k}}) $  сферическая функция от единичного  трехмерного вектора $ \hat{\mathbf{k}}=\dfrac{\mathbf{k}}{k} $.
\begin{multline} \label{2.3}
\eta_{n,l}(k)=
4^{l+1}l!\, \sqrt{\frac{n!\,(n{+}l{+}1)}{\pi\,(n{+}2l{+}1)!}} \frac{k^l}{\left(k^2{+}1\right)^{l+\tfrac{3}{2}}}
C_n^{l+1}\left( \frac{k^2{-}1}{k^2{+}1} \right)=
\\
=\frac{2}{\Gamma\left(l{+}\dfrac{3}{2}\right)}
\sqrt{\frac{(n{+}l{+}1)\,(n{+}2l{+}1)!}{n!}} 
\frac{k^l}{\left(k^2{+}1\right)^{l+\frac{3}{2}}}
\,{}_2F_1 \left[ \left. 
\begin{matrix}
{  -n  \quad n{+}2l{+}2  } \\
{  l{+}\dfrac{3}{2}  }
\end{matrix} \right| \frac{1}{k^2{+}1}\right] 
\end{multline}
Здесь и далее  $ \,{}_{2}F_{1} \left[ \left. \ldots \right|  z  \right] $-- гипергеометрическая функция Гаусса,  $ (a)_n=\dfrac{\Gamma( a+n )}{\Gamma( a )} $--символ Похгаммера. Отметим, что  $ \eta_{n,l}(k) $-функции ортонормированные с весом $ k^2 $  на всей действительной оси $ k\geqslant 0 $ , и соответственно ортогональны \eqref{2.2} ( $ d\mathbf{k}=k^2\,dk\,d\Omega_{\hat{\mathbf{k}}} $--элемент объема,  $ d\Omega_{\hat{\mathbf{k}}} $--элемент телесного угла)
\begin{equation} \label{2.4}
\int\!\! d\mathbf{k}\, 
H_{n_1,l_1,m_1}\left( \frac{\mathbf{k}}{\gamma} \right)
H_{n_2,l_2,m_2}^{\ast}\left( \frac{\mathbf{k}}{\gamma} \right)=
\gamma^3 \delta_{n_1,n_2}\delta_{l_1,l_2}\delta_{m_1,m_2}
\end{equation}
Здесь $ \gamma \in\mathbb{R} $  . Очевидно, что из вида представления \eqref{2.1}, решением интегрального уравнения \eqref{1.1} можно представить в виде как
\begin{multline} \label{2.5}
\left< \mathbf{k}_2 \left|T(z)  \right| \mathbf{k}_1 \right>=
\frac{\alpha}{2\gamma^{4}}(2\pi)^{\frac{3}{2}}\,
\sqrt{k_2^2+\gamma^2}\sqrt{k_1^2+\gamma^2}
\times
\\
\times
\sum_{n_2,n_1,l,m}\!\!\!\!\sqrt{\upsilon_{n_2,l}\upsilon_{n_1,l}}\tau_{n_2,n_1;l}(z)
H_{n_2,l,m}\left( \frac{\mathbf{k}_2}{\gamma} \right)
H_{n_1,l,m}^{\ast}\left( \frac{\mathbf{k}_1}{\gamma} \right)
\end{multline}
где элементы   $ \tau_{n_2,n_1;l}(z) $ из условия \eqref{2.4} определяются системой алгебраических уравнений.
\begin{equation} \label{2.6}
\tau_{n_2,n_1;l}(z)=\sigma\delta_{n_2,n_1}+\sigma\sum_{n_3=0}^{\infty}A_{n_2,n_3;l}^{-1}\tau_{n_3,n_1;l}(z)
\end{equation}
где
\begin{equation}  \label{2.7}
A_{n_2,n_3;l}^{-1}=\frac{\alpha\mu}{\gamma\hbar^2}\, \sqrt{\upsilon_{n_2,l} \upsilon_{n_3,l}}
\int\limits_{0}^{\infty}\!\! dx\, x^2\,\frac{x^2+1}{\dfrac{2\mu z}{\gamma^2\hbar^2}-x^2}\eta_{n_2,l}(x)\eta_{n_3,l}(x)
\end{equation}
или в одной из  матричной формы
\begin{gather} 
\boldsymbol{\tau}=\sigma\left( \mathbf{E}-\sigma\mathbf{A}^{-1} \right)^{-1} \notag
\\
\label{2.8}
\boldsymbol{\tau}=\sigma \left( \mathbf{A}-\sigma\mathbf{E} \right)^{-1}\!\mathbf{A}
\end{gather}
($ \mathbf{E} $--единичная матрица). Из полноты ортогональных функции, элементы матрицы $ \mathbf{A} $  представляются в виде как
\begin{equation} \label{2.9}
A_{n_1,n_2;l}=\!\frac{\gamma\hbar^2}{\alpha\mu} \frac{1}{\sqrt{\upsilon_{n_1,l}\upsilon_{n_2,l}}}
\int\limits_0^\infty \!\! dx\,x^2\, \frac{\dfrac{2\mu z}{\gamma^2\hbar^2}-x^2}{x^2+1}\eta_{n_1,l}(x)\eta_{n_2,l}(x)
\end{equation}
Используя рекуррентное соотношение для гипергеометрической функции Гаусса \cite[Гл.2]{Beitman_1}  в \eqref{2.3}, где верхние индексы отличаются на единицу, можно получить рекуррентное соотношение для функции $ \eta_{n,l}(x) $  в виде
\begin{equation*}
\frac{x^2-1}{x^2+1}\eta_{n,l}(x)=
\frac{1}{2}\sqrt{\frac{(n+1)(n+2l+2)}{(n+l+1)(n+l+2)}}\eta_{n+1,l}(x)+
\frac{1}{2}\sqrt{\frac{n(n+2l+1)}{(n+l)(n+l+1)}}\eta_{n-1,l}(x)
\end{equation*}
учитывая условие ортогональности  $ \eta_{n,l}(x) $-функции, а также вводя вспомогательные элементы
\begin{equation} \label{2.10}
\phi_n=i\sqrt{\frac{y+1}{2\rho}}
\sqrt{\frac{\Gamma\left( \dfrac{n}{2}+1 \right)\Gamma\left( \dfrac{n+3}{2}+l \right)}{\Gamma\left( \dfrac{n+1}{2} \right)\Gamma\left( \dfrac{n}{2}+l+1 \right)}}
\end{equation}
где
\begin{equation} \label{UNC_4}
y=\frac{2\mu z}{\gamma^2 \hbar^2}, \quad \rho=\frac{\alpha\mu}{\gamma\,\hbar^2}
\end{equation}
представим \eqref{2.9} в простом матричном виде
\begin{equation} \label{2.11}
A_{n_1,n_2}=\frac{y-1}{2\rho} (n_1+l+1)\delta_{n_1,n_2}+\phi_{n_1}\phi_{n_2}(\delta_{n_1+1,n_2}+\delta_{n_1-1,n_2})
\end{equation}
Для введения вспомогательного элемента $ \phi_n $ мы использовали соотношение вида \cite[Гл.1, п.1.2]{Beitman_1} 
\begin{equation*}
\Gamma\left( \frac{n}{2}+z \right) \Gamma\left( \frac{n+1}{2}+z \right)=2^{-2z-n+1}\,\sqrt{\pi}\,\Gamma( n+2z )
\end{equation*}
Отметим, что матрица $ \mathbf{A}^{-1} $  в \eqref{2.7}, как и $ \mathbf{A} $  в \eqref{2.11} – бесконечномерные. Интеграл в \eqref{2.7} сходится всегда и сходится даже для потенциалов $ \sim r^{-v} $  при $ v<\dfrac{5}{2} $ . Однако его значение сильно зависит от $ z $ , и может принимать разный вид как для действительных положительных, действительных отрицательных и комплексных значении $ z $ . Хотя мы всегда можем выбрать такой параметр $ \gamma $  для однозначного вида интеграла, но при комплексном значении $ \gamma $  теряется свойство ортогональности \eqref{2.4}, что крайне не желательно в дальнейшем, и поэтому лучше придерживаться условием $ \gamma^2>0 $ . Преимущество \eqref{2.9} перед \eqref{2.7} заключается в том, что $ z $ может входить входит как параметр и не влиять на интегралы, что и было получено \eqref{2.11}.

В самом простом частном случае, когда $ z=-E \in \mathbb{R} $  , и выбирая параметр  $ \gamma^2=\dfrac{2\mu E}{\hbar^2} $ получим, что матричные элементы \eqref{2.7} (как и \eqref{2.11}) диагональные а элементы \eqref{2.6} представятся в простом виде как
\begin{equation} \label{UNC_5}
\tau_{n_2,n_1;l}(-E)=\delta_{n_2,n_1}\sigma\left( 1+\sigma\,\sqrt{\frac{E_{n_1,l}}{E}} \right)^{-1},\quad E_{n_1,l}=\frac{E_b}{(n_1+l+1)^2}
\end{equation}
где  $ E_b=13.6 $ эВ-энергия основного состояния. При $ \sigma{=}-1 $ , $ \tau_{n_2,n_1;l}(-E) $   имеет полюс $ E=E_{n_1,l} $  с вычетом $  -2\, \sqrt{\left( E_{n_1,l} \right)^3}  $.

Теперь определим $ \boldsymbol{\tau} $  из \eqref{2.8} при любом  $ z $ (принимающие комплексные или действительные значения) и произвольным параметром $ \gamma\in\mathbb{R} $ . Из \eqref{2.11} видим, что $ \mathbf{A} $   симметрична и трёхдиагональная. Поэтому представим ее как в виде произведения от двух бесконечномерных верхнеугольных матриц
\begin{equation} \label{2.12}
\mathbf{A}=\mathbf{b}\mathbf{b}^{T}
\end{equation}
где элементы $ \mathbf{b} $  представим в виде
\begin{equation*} 
b_{n,m}=\phi_n\sqrt{\beta_m}\,\delta_{n+1,m}+\frac{\phi_n}{\sqrt{\beta_m}}\,\delta_{n,m}
\end{equation*}
Данное представление \eqref{2.12} будет справедливым, если $ \beta_n $  будут удовлетворять рекуррентному соотношению вида
\begin{equation} \label{2.13}
\phi_n^2 \left( \beta_{n+1}+\frac{1}{\beta_n} \right)=
\frac{y-1}{2\rho}\, (n+l+1)-\sigma=\lambda_n,\,\, n=0,1,2,\ldots
\end{equation}
Таким образом \eqref{2.8} можно представить в виде как
\begin{equation} \label{2.14}
\boldsymbol{\tau}=\sigma \mathbf{B}^{T}\mathbf{B}\mathbf{A}
\end{equation}
где $ \mathbf{B}=\mathbf{b}^{-1} $  и его элементы имеют вид
\begin{equation} \label{2.15}
B_{n,m}=\begin{cases}
\dfrac{(-1)^{n+m}}{\phi_m\sqrt{\beta_n}}\prod\limits_{i=n}^{m}\beta_i,& m\geqslant n \\
0, & m<n
\end{cases}
\end{equation}
и соответственно легко получить
\begin{equation} \label{UNC_1}
(\mathbf{A}-\sigma\mathbf{E})^{-1}_{n,m}= \mathbf{B}^{T}\mathbf{B}=
\begin{cases}
\dfrac{(-1)^{n+m}}{\phi_n\phi_m}\left( 1+\sum\limits_{i=0}^{m-1}\prod\limits_{j=0}^{i}\beta_{m-j-1}\beta_{m-j} \right)\prod\limits_{k=m}^{n}\beta_k, & n\geqslant m \\
\text{то же самое при перестановки } n\leftrightarrow m, & n<m
\end{cases}
\end{equation}
Представление \eqref{2.14} не очень практично, даже в случае конечномерных матриц. Поэтому сделаем следующим образом. Представим в \eqref{2.14} произведение
\begin{equation} \label{2.16}
\mathbf{B}\mathbf{A}=\mathbf{S}\mathbf{B}
\end{equation}
где элементы $ \mathbf{S} $  есть как
\begin{equation*}
S_{n,m}=\left( \sigma+\frac{\phi_n^2}{\beta_n}+\beta_n \phi_{n-1}^2 \right)\delta_{n,m}+\phi_n^2\,\sqrt{\frac{\beta_m}{\beta_n}}\,\delta_{n+1,m}+\phi_m^2\,\sqrt{\frac{\beta_n}{\beta_m}}\,\delta_{n,m+1}
\end{equation*} 
Так как   $ \mathbf{S} $--симметрична, разобьем его в виде как
\begin{equation} \label{2.17}
\mathbf{S}=\mathbf{s}^{T}\mathbf{G}\,\mathbf{s}
\end{equation}
где 
\begin{equation*}
G_{n,m}=\frac{g_n}{1-g_n}\,\delta_{n,m}
\end{equation*}
\begin{equation} \label{2.18}
s_{n,m}=\frac{\phi_n}{\sqrt{\beta_n}}\,\delta_{n,m}+\frac{1-g_n}{g_n}\,\phi_n\,\sqrt{\beta_m}\,\delta_{n+1,m}
\end{equation}
Объединяя \eqref{2.17}, \eqref{2.16} с \eqref{2.14}, запишем
\begin{equation*}
\boldsymbol{\tau}=\sigma\, \mathbf{C}^{T}\mathbf{G}\,\mathbf{C}
\end{equation*}
где $ \mathbf{C}=\mathbf{s}\mathbf{B} $ , а из \eqref{2.18}, и \eqref{2.15} его элементы будут иметь вид
\begin{equation} \label{2.19}
C_{n,m}=
\begin{cases}
(-1)^{n+m}\, \dfrac{\phi_n}{\phi_m}\, \dfrac{2g_n-1}{g_n}\prod\limits_{k=n+1}^{m}\beta_k, & m>n 
\\
1, & m=n
\\
0, & m<n
\end{cases}
\end{equation}
Отметим, что в разбиении \eqref{2.17} $ g_n $   должны удовлетворять рекуррентным соотношением вида
\begin{subequations} \label{2.20}
	\begin{equation} \label{2.20a}
	\frac{2g_{n+1}-1}{1-g_{n+1}}\,\frac{\phi^2_{n+1}}{\beta_{n+1}}-\frac{2g_n-1}{g_n}\,\beta_{n+1} \phi^2_n-\sigma=0,\,\,n=0,1,2,\ldots
	\end{equation}
	с начальным условием 
	\begin{equation} \label{2.20b}
	g_0=\frac{\phi_0^2+\sigma\beta_0}{2\phi_0^2+\sigma\beta_0}
	\end{equation}
\end{subequations}
Введем новые функции вида
\begin{subequations} \label{2.21}
	\begin{equation} \label{2.21a}
	\Phi_{n,l,m}(\mathbf{k},z)=\sqrt{\frac{\alpha}{2}}\,\frac{\sqrt{k^2+\gamma^2}}{\gamma^2}
	\sum_{n_1=n}^{\infty}C_{n,n_1}\,\sqrt{\upsilon_{_{n_1,l}}}H_{n_1,l,m}\left( \frac{\mathbf{k}}{\gamma} \right)
	\end{equation}
	тогда \eqref{2.5} можно представить в диагональном виде как
	\begin{equation*}
	\left< \mathbf{k}_2 \left|T(z) \right| \mathbf{k}_1 \right>=-(2\pi)^{\frac{3}{2}}\,\sigma \sum_{n,l,m} \frac{g_n}{1-g_n} \hat{\Phi}_{n,l,m}(\mathbf{k}_1,z^{\ast})\Phi_{n,l,m}(\mathbf{k}_2,z)
	\end{equation*}
	где под $ \hat{\Phi}_{n,l,m} $  подразумевается транспонирование и комплексное сопряжение, или в виде 
	\begin{equation} \label{2.21b}
	\hat{\Phi}_{n,l,m}(\mathbf{k},z^{\ast})=\sqrt{\frac{\alpha}{2}}\,\frac{\sqrt{k^2+\gamma^2}}{\gamma^2}
	\sum_{n_1=n}^{\infty}C_{n,n_1}\,\sqrt{\upsilon_{_{n_1,l}}}H^{\ast}_{n_1,l,m}\left( \frac{\mathbf{k}}{\gamma} \right)
	\end{equation}
\end{subequations}
Отметим, что при $ \sigma{=}0 $  из \eqref{2.20} все $ g_n{=}\dfrac{1}{2} $  , и элементы $ \mathbf{C} $ , как видно из \eqref{2.19} будут единичными, что как и следовало бы ожидать из \eqref{2.8}.

\section{Определение $ \beta_n $.}

Из \eqref{2.13} видим, что общее решение для $ \beta_n $  определяется в виде бесконечной цепной дроби, если начальное значение $ \beta_0 $  не задано. Или в виде конечной цепной дроби, если $ \beta_0 $ задано. Здесь мы рассмотрим второй случай.

Представим $ \beta_n $  в виде
\begin{subequations} \label{3.1}
	\begin{equation}  \label{3.1a}
	\beta_n=\phi_n^2\,\frac{2\rho}{y+1}\,\frac{2}{n+2l+1}\,\frac{R_n}{R_{n-1}}
	=\phi_n^2\,\frac{2\rho}{y+1}\,\frac{\Gamma\left( \dfrac{n+1}{2}+l \right)}{\Gamma\left( \dfrac{n+3}{2}+l \right)}\,\frac{R_n}{R_{n-1}},\quad (n=0,1,2,\ldots)
	\end{equation}
	Подставляя сюда \eqref{2.10} мы также можем представить в другом виде
	\begin{equation}  \label{3.1b}
	\beta_n=\frac{Q_n}{Q_{n-1}},\,\,Q_n=(-1)^n\,R_n\,\frac{\Gamma\left( \dfrac{n+2}{2} \right)}{\Gamma\left( \dfrac{n+2}{2}+l \right)}=\frac{(-1)^n\,R_n}{\left( \dfrac{n+2}{2} \right)_l}
	\end{equation}
\end{subequations}
Данные представления удобны тем, что $ \prod\limits_{k=n}^{m}\beta_k=\dfrac{Q_m}{Q_{n-1}} $ . Так, из \eqref{2.19} получим для функции \eqref{2.21a} простой вид
\begin{multline} \label{3.2}
\Phi_{n,l,m}(\mathbf{k},z)=
\sqrt{\frac{\alpha}{2}}\,\frac{\sqrt{k^2+\gamma^2}}{\gamma^2} 
\sqrt{\upsilon_{_{n,l}}}H_{n,l,m}\left( \frac{\mathbf{k}}{\gamma} \right)+
\\
+(-1)^n\, \sqrt{\frac{\alpha}{2}}\,\frac{\sqrt{k^2+\gamma^2}}{\gamma^2}\,\frac{\phi_n}{Q_n}\,\frac{2g_n-1}{g_n}  \sum_{n_1=n+1}^{\infty}(-1)^{n_1}\,\frac{Q_{n_1}}{\phi_{n_1}}\,
\sqrt{\upsilon_{_{n_1,l}}}H_{n_1,l,m}\left( \frac{\mathbf{k}}{\gamma} \right)
\end{multline}
и аналогично для \eqref{2.21b}. \\
Из \eqref{3.1a} и \eqref{2.13} получим рекуррентное соотношения для $ R_n $   
\begin{equation} \label{3.3}
(n+1)R_{n+1}-2x\left( n+l+1-\frac{2\sigma\rho}{y-1} \right)R_n+(n+2l+1)R_{n-1}=0,\,\,(n=0,1,2,\ldots)
\end{equation}
где
\begin{equation*}
x=\frac{y-1}{y+1}
\end{equation*}
с начальными условиям при $ R_{-1}\neq 0 $  и $ R_0\neq 0 $ , так как из \eqref{3.1b} $ \beta_0 $   должно быть регулярным. \\ 
Производящую функцию для $ R_n $  
\begin{equation*}
g(\omega)=\sum_{n=0}^{\infty}R_n\,\omega^n
\end{equation*} 
можно получить методами как в \cite{Jones_Thron} , которая будет соответствовать виду 
\begin{multline*}
g(\omega)=R_0 \left( 1-\frac{\omega}{\omega_2^{\ast}} \right)^{-l-1+i\frac{\sigma\rho}{\sqrt{y}}} \left( 1-\frac{\omega}{\omega_2} \right)^{-l-1-i\frac{\sigma\rho}{\sqrt{y}}}-
\\
-R_{-1} \left( 1-\frac{\omega}{\omega_2^{\ast}} \right)^{-l-1+i\frac{\sigma\rho}{\sqrt{y}}} \left( 1-\frac{\omega}{\omega_2} \right)^{-l-1-i\frac{\sigma\rho}{\sqrt{y}}}
\int\limits_{0}^{\omega}\!\! d \xi  \,\left( 1-\frac{\xi}{\omega_2^{\ast}} \right)^{l-i\frac{\sigma\rho}{\sqrt{y}}} \left( 1-\frac{\xi}{\omega_2} \right)^{l+i\frac{\sigma\rho}{\sqrt{y}}}
\end{multline*}
где при любых комплексных значении $ y $
\begin{equation*}
\omega_2=\frac{y-1}{y+1}-i\,\frac{2\,\sqrt{y}}{y+1}
\end{equation*}
Из \cite{Pollaczec_Acad_1950_230_1563,Pollaczec_Acad_1950_230_2254} можно получить общее выражение для $ R_{-1},R_0,R_n,(n=1,2,\ldots) $  при любых комплексных (и положительных действительных) чисел $ y $
\begin{equation} \label{3.4}
R_n=R_{-1}\,\frac{\omega_2^{n+1}\,(2l+1)_{n+1}}{\left( l+1+i\,\dfrac{\rho\sigma}{\sqrt{y}} \right)_{n+1}}
\,{}_{2}F_{1} \left[ 
\left.
\begin{gathered}
{ n+1 \quad -l+i\,\frac{\rho\sigma}{\sqrt{y}} } \\
{ n+l+2+i\,\frac{\rho\sigma}{\sqrt{y}} }
\end{gathered}
\right|  \omega_2^2  \right]
\end{equation}
и когда $ y $ принимает действительные отрицательные значения $ y=-t,\,\,t>0 $ 
\begin{equation} \label{3.5}
R_n=R_{-1}\left( \frac{\sqrt{t}-1}{\sqrt{t}+1} \right)^{n+1} \frac{(2l+1)_{n+1}}{\left( l+1+\dfrac{\rho\sigma}{\sqrt{t}} \right)_{n+1}}
\,{}_{2}F_{1} \left[ 
\left.
\begin{gathered}
{ n+1 \quad -l+\frac{\rho\sigma}{\sqrt{t}} } \\
{ n+l+2+\frac{\rho\sigma}{\sqrt{t}} }
\end{gathered}
\right|  \left( \frac{\sqrt{t}-1}{\sqrt{t}+1} \right)^2  \right] 
\end{equation}
Отметим, что $ R_n $  в \eqref{3.4} комплексное а $ R_{-1} $  задается из \eqref{3.1b} при заданном начальным значением $ \beta_0 $ . Так как в \eqref{3.1} и \eqref{3.2} $ Q_n $   (как и $ R_n $ ) входят в виде отношения, то во всех приведенных формулах $ \beta_n $  и  $ \Phi_{n,l,m}(\mathbf{k}) $ уже не зависит от начального заданного $ \beta_0 $ . В виду линейности рекуррентного соотношения \eqref{3.3}, кроме общего выражения \eqref{3.4} при $ \Im y=0 $  мы можем взять по отдельности как действительные так и мнимые части. Соответственно представим асимптотическое поведение в сумме \eqref{3.2} при больших $ n_1 $. Для \eqref{3.4} (при $ \Im y=0 $)
\begin{multline*}
(-1)^{n_1}\,\frac{Q_{n_1}}{\phi_{n_1}}\,\upsilon_{_{n_1,l}}=
\\
=R_{-1}\,\frac{2^{l+\frac{1}{2}}\Gamma\left( l+1+i\,\dfrac{\rho\sigma}{\sqrt{y}} \right)}{\Gamma( 2l+1 )}\,e^{-i\left( \theta(n_1+1)+\frac{\rho\sigma}{\sqrt{y}}\ln n_1 \right)}
\left( 1-e^{-i\,2\theta} \right)^{l-i\,\frac{\rho\sigma}{\sqrt{y}}}\frac{1}{n_1}\left( 1+O\left( \frac{1}{n_1} \right) \right)
\end{multline*}
где $ \cos\theta{=}\dfrac{y-1}{y+1} $. И для \eqref{3.5}
\begin{multline*}
(-1)^{n_1}\,\frac{Q_{n_1}}{\phi_{n_1}}\,\upsilon_{_{n_1,l}}=
\\
=R_{-1}\,\frac{2^{l+\frac{1}{2}}\Gamma\left( l+1+\dfrac{\rho\sigma}{\sqrt{t}} \right)}{\Gamma( 2l+1 )}\,\left( \frac{\sqrt{t}-1}{\sqrt{t}+1} \right)^{n_1+1} n_1^{-\frac{\rho\sigma}{\sqrt{t}}} \left( \frac{4\sqrt{t}}{(\sqrt{t}+1)^2} \right)^{l-\frac{\rho\sigma}{\sqrt{t}}}
\frac{1}{n_1}\left( 1+O\left( \frac{1}{n_1} \right) \right)
\end{multline*}
Так как для быстро осциллирующей функции $ e^{-i\left( \theta(n_1+1)+\frac{\rho\sigma}{\sqrt{y}}\ln n_1 \right)}<1 $  то коэффициент в первом выражении будет убывать как $ \dfrac{1}{n_1} $ . Для второго выражения сходимость в \eqref{3.2} будет быстрее только для отталкивательного потенциала ($ \sigma{=}+1 $).

\printbibliography[heading=bibintoc]
\end{document}